\documentclass[12pt]{article}
\usepackage{geometry}	
\usepackage{graphicx} 
\usepackage{epstopdf}
\usepackage{url}	
\usepackage{setspace}\spacing{1.213}	
\usepackage[bottom]{footmisc}			
\usepackage{rotating,float}		
\usepackage[font=small]{caption} 
\usepackage{subcaption}
\usepackage{verbatim}  
\usepackage{adjustbox} 
\usepackage{relsize}   
\usepackage{amsmath,amssymb}\allowdisplaybreaks[1] 

\usepackage{appendix} 
\usepackage{placeins}
\usepackage[small,compact]{titlesec}  
\usepackage[bookmarks=true]{hyperref} 
\usepackage{bookmark}
\usepackage{pdflscape}
\usepackage{url}

\usepackage{xcolor}	
\hypersetup{
	colorlinks,
	linkcolor={red!50!black},
	citecolor={blue!50!black},
	urlcolor={blue!80!black}
}

\usepackage[authoryear]{natbib}

\usepackage{amsthm}
\theoremstyle{plain}

\theoremstyle{definition}

\newtheoremstyle{indenteddefinition}
	{}
	{}
	{\hangindent=2em}
	{}
	{\bfseries}
	{.}
	{.5em}
	{}
\theoremstyle{indenteddefinition}

\newcounter{assumptiongroup}\stepcounter{assumptiongroup}

\title{Counting Defiers\thanks{This manuscript was first circulated as NBER Working Paper 25671 in March of 2019. I thank Neil Christy, Tory Do, Bailey Flanigan, Anastasiya Karpova, Pauline Mourot, and Matthew Tauzer for excellent research assistance.  I thank Charles Antonelli, Bennett Fauber, and Advanced Research Computing at the University of Michigan for computational support.  I also thank my parents and my sister for their support.  I have never thought of us a family of defiers, but we could be.}}
\author{Amanda E. Kowalski}
\date{March 13, 2019}

\begin{document}
\maketitle

\begin{center}
This paper has been combined with ``A Model of a Randomized Experiment with an Application to the PROWESS Clinical Trial'' (\url{https://arxiv.org/abs/1908.05810}) and superseded by ``Counting Defiers: Examples from Health Care'' (\url{https://arxiv.org/abs/1912.06739}) as of 

July 17, 2020.
\end{center}
\bigskip	

\begin{abstract}
\noindent The LATE monotonicity assumption of \citet{imbens1994} precludes ``defiers,'' individuals whose treatment always runs counter to the instrument, in the terminology of \citet{balke1993} and \citet{angrist1996}.   I allow for defiers in a model with a binary instrument and a binary treatment.  The model is explicit about the randomization process that gives rise to the instrument.  I use the model to develop estimators of the counts of defiers, always takers, compliers, and never takers.  I propose separate versions of the estimators for contexts in which the parameter of the randomization process is unspecified, which I intend for use with natural experiments with virtual random assignment.  I present an empirical application that revisits \citet{angrist1998}, which examines the impact of virtual random assignment of the sex of the first two children on subsequent fertility.  I find that subsequent fertility is much more responsive to the sex mix of the first two children when defiers are allowed.  
\end{abstract}

\newpage
\section{Introduction}

Consider a model with a binary treatment and a binary instrument.  In the terminology of \citet{balke1993} and \citet{angrist1996}, ``defiers'' are individuals whose treatment always runs counter to the instrument.  The LATE monotonicity assumption of \citet{imbens1994} precludes defiers. However, it seems plausible that defiers could be possible in some contexts.     
    
For example, consider the context of \citet{angrist1998}, in which the instrument is an indicator for whether an individual's first two children are of the same sex. The treatment is an indicator for whether the individual has a third child.  Having a third child is a ``treatment'' in the sense that the authors are interested in the ``treatment effect'' of having a third child on labor supply.  

In the  \citet{angrist1998} context, defiers are individuals that would have a third child if and only if their first two children are of different sex.  It seems plausible that such individuals could exist.  For example, suppose that some individuals want a big family, but they want to avoid having three sons.  Such individuals are defiers: they would have a third child if their first two children were a son and a daughter, but they would not have a third child if their first two children were sons.

Even though it seems plausible that defiers could exist, it is difficult to identify any individual as a defier.  This difficulty has been formalized with the concept of ``potential outcomes'' \citep{rubin1974,rubin1977,holland1986}.   In a model with a binary treatment and a binary instrument, each individual has two potential outcomes for the treatment, one for each value of the instrument, so there are four possible types of individuals.  Defiers are one type, whose treatment always runs counter to the instrument.  The terminology of \citet{angrist1996} also includes ``compliers,'' whose treatment always aligns with the instrument, ``always takers" who are treated regardless of the value of the instrument, and ``never takers'' who are untreated regardless of the value of the instrument. For each individual, only one value of the instrument is realized, so only one potential outcome is observed, which makes it difficult to identify any individual as a member of any one of the four groups. Returning to the \citet{angrist1998} context, an individual with two sons and no third child could be a defier or a never taker, depending on whether the individual would have had a third child if the first two children had been of the same sex.

I propose estimators of the counts of defiers, always takers, compliers, and never takers in a model with a binary instrument and a binary treatment.  The model eliminates the LATE montonicity assumption but maintains the LATE independence assumption of \citet{imbens1994}.  Random assignment of the instrument satisfies the the LATE independence assumption.  Accordingly, I use the model to be explicit about the randomization process that gives rise to the instrument.  To do so, I employ the language of a randomized experiment in which the instrument is a binary variable that indicates assignment to the intervention group as opposed to the control group.  

A key feature of the model is that the randomization process results in randomization error.  The randomization process is a Bernoulli process with a parameter that represents the intended fraction of individuals in the intervention group.  The randomization error is the difference between the actual and intended numbers of individuals in the intervention group.  Using the distribution of the randomization error for defiers, compliers, always takers, and never takers, I derive the probability of the observed data.  

The observed data consist of four numbers.  These numbers include the number of treated individuals assigned to the intervention group, the number of untreated individuals assigned to the intervention group, the number of treated individuals assigned to the control group, and the number of untreated individuals assigned to the control group.  I use the model to derive the probability of the observed data as a function of the counts of defiers, compliers, always takers, and never takers, given the intended fraction of individuals in the intervention group.  I use the probability of the observed data to specify the likelihood of the vector of counts of defiers, compliers, always takers, and never takers, given the intended fraction of individuals in the intervention group. 

The first estimator is a maximum likelihood estimator that maximizes the probability of the observed data, given the intended fraction of individuals in the intervention group.  This estimator is not yet computationally tractable via a variety of approaches.  Therefore, I also propose a second estimator. The second estimator is a least squares estimator that minimizes a weighted sum of the randomization error within each type, given the observed data and the intended fraction of individuals in the intervention group.

This paper is most closely related to my work in \citet{kowalski2019}.  In that paper, I present a reduced form model with a binary instrument and a binary outcome.  Here, I focus on a first stage model with a binary instrument and a binary treatment.  The LATE framework of \citet{imbens1994} imposes monotonicity in the first stage but not in the reduced form.  The main contribution of this paper is that I adapt the model and estimators from \citet{kowalski2019} to the first stage.  In doing so, I illustrate that there is a fundamental isomorphism between the reduced form and the first stage that allows me to eliminate the LATE monotonicity assumption in the first stage.

A second contribution of this paper is that I present separate versions of the estimators for contexts in which the intended fraction of individuals in the intervention group is unspecified. It might be desirable to specify this parameter in some contexts but not in others.  For example, if the protocol of a randomized experiment specifies an intended fraction of individuals in the intervention group, it might be desirable to specify it in estimation. However, in a context of a natural experiment with virtual random assignment such as \citet{angrist1998}, it is unclear what the intended fraction of individuals in the intervention group should be, so it might be desirable to leave it unspecified in estimation.  Accordingly, I propose a separate version of each estimator that incorporates the intended fraction of individuals in the intervention group as an additional parameter to be estimated.  In doing so, I adapt the maximum likelihood and least squares estimators for use with natural experiments.    

A third contribution of this paper is that I present an empirical application to \citet{angrist1998}, a natural experiment in which defiers seem plausible.  The first stage estimate shows that individuals whose first two children are of the same sex are approximately 6 percentage points more likely to have a third child than individuals whose first two children are of opposite sex.  Under the LATE monotonicity assumption, the first stage estimate gives the share of compliers, so it implies that compliers make up approximately 6\% of the sample.  Without the LATE monotonicity assumption, the first stage estimate gives the share of compliers minus the share of defiers \citep{angrist1996}.  Using the least squares estimator, I find that compliers make up approximately 25\% of the sample and defiers make up approximately 19\% of the sample.  Subsequent fertility responds to the sex mix of the first two children for compliers and defiers.  Given the large share of defiers, subsequent fertility responds much more to the sex mix of the first two children when defiers are allowed.

I present the model in the next section, and I discuss estimation in Section~\ref{sec:estimation}.  I present the empirical application to \citet{angrist1998} in Section~\ref{sec:empirical}.  I discuss implications and conclude in Sections~\ref{sec:discussion} and \ref{sec:conclusion}.    

\section{Model}\label{sec:model}

Let $Z$ represent a binary instrument.  Using the language of a randomized experiment, $Z$ represents assignment to the intervention group, such that participants with $Z=1$ are in the intervention group, and participants with $Z=0$ are in the control group. Let $D$ represent a binary treatment such that participants with $D=1$ are treated and participants with $D=0$ are untreated.        

\begin{figure}[!htb]
	\caption{Matrix that Relates Potential Outcome Types and Observed Outcome Groups}
	\centering	
	\includegraphics[width=\textwidth]{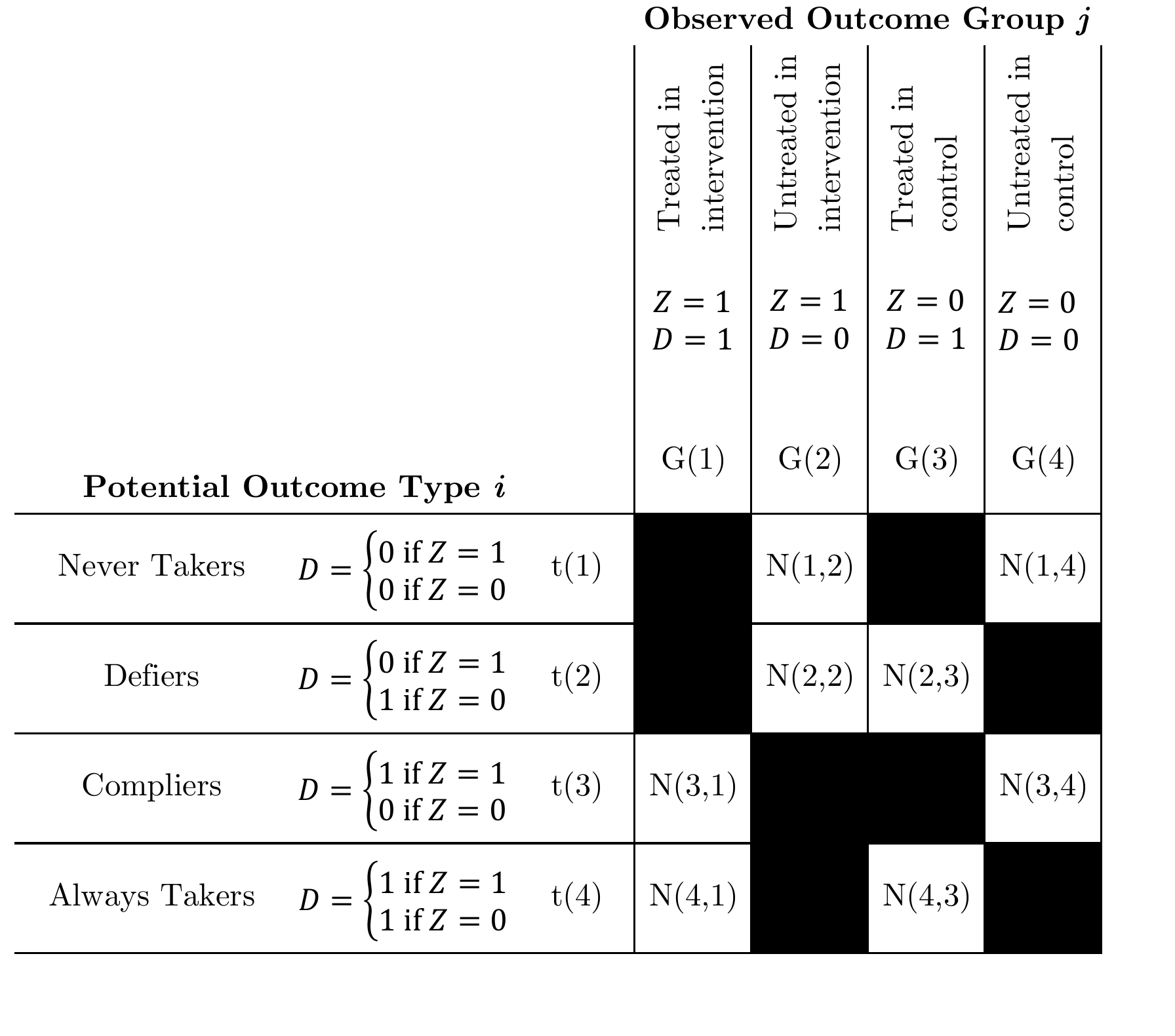}
	\label{fig:matrix} \\
    \vspace{-40pt}
	\begin{minipage}{1\linewidth}
	\scriptsize
	\emph{Note}. $D$ is the treatment, and $Z$ is the instrument, which represents assignment to the intervention group. $N(i,j)$, the number of participants of potential outcome type $i$ in observed outcome group $j$, must be equal to zero in all shaded cells.  The intended fraction of participants in the intervention group is $p$.
\end{minipage}
\end{figure}

Each participant has a potential outcome type $i$.  There are four possible potential outcome types, as shown in the rows of the matrix in Figure~\ref{fig:matrix}.  Never takers, type $i=1$, are untreated regardless of assignment to the intervention or control group: $D=0$ if $Z=1$ and $D=0$ if $Z=0$.  Defiers,  type $i=2$, are untreated when assigned to the intervention group and treated when assigned to the control group: $D=0$ if $Z=1$ and $D=1$ if $Z=0$.  Compliers, type $i=3$, are treated when assigned to the intervention group and untreated when assigned to the control group: $D=1$ if $Z=1$ and $D=0$ if $Z=0$. Finally, always takers, type $i=4$, are treated regardless of assignment to the intervention or control group: $D=1$ if $Z=1$ and $D=1$ if $Z=0$.    

After random assignment, each participant also belongs to an observed outcome group $j$.  There are four possible observed outcome groups, as shown in the columns of the matrix in Figure~\ref{fig:matrix}.  Group $j=1$ includes treated individuals in the intervention group ($Z=1$ and $D=1$).  Group $j=2$ includes untreated individuals in the intervention group ($Z=1$ and $D=0$). Group $j=3$ includes treated individuals in the control group ($Z=0$ and $D=1$).  Finally, group $j=4$, includes untreated individuals in the control group ($Z=0$ and $D=0$).

Random assignment determines the observed outcome group $j$ for each participant of potential outcome type $i$.  For example, consider a participant that is a never taker, type $i=1$.  Never takers are untreated regardless of assignment to the intervention group.  Therefore, a never taker assigned to the intervention group will belong to group $j=2$, which includes the untreated participants in the intervention group.  Alternatively, a never taker assigned to the control group will belong to group $j=4$, which includes the untreated participants in the control group.  A never taker will never belong to group $j=1$ or group $j=3$ because participants in both of those groups are treated. As if solving a logic puzzle, I shade the cells of the matrix in Figure~\ref{fig:matrix} that represent combinations of potential outcome types and observed outcome groups that cannot occur.  

Random assignment follows a Bernoulli process with parameter $p$. In some contexts, the experimental protocol specifies $p$ as the intended fraction of participants in the intervention group.  In other contexts, such as natural experiments, nature determines $p$.  Regardless, $p$ is the same for all participants.  Therefore, the indicator for assignment to the intervention group is an independent an identically distributed Bernoulli random variable with parameter $p$.    

In each row of the matrix in Figure~\ref{fig:matrix}, let $t(i)$ denote the number of participants of type $i$.  The number of participants of type $i$ who are assigned to the intervention group is a binomial random variable with parameters $t(i)$ and $p$.  This follows because the sum of independent and identically distributed Bernoulli random variables has a binomial distribution with parameters that represent the number of terms in the sum and the parameter of the Bernoulli distribution.  

In each cell of the matrix in Figure~\ref{fig:matrix}, let $N(i,j)$ denote the number of participants of type $i$ in group $j$, where I use a capital letter to emphasize that it is a random variable.  Per the matrix in Figure~\ref{fig:matrix}, the number of participants of type $i$ who are assigned to the intervention group is equal to $N(i,1)$+$N(1,2)$.  Therefore, number of participants of type $i$ who are assigned to the intervention group, $N(i,1)$+$N(i,2)$ is a binomial random variable with parameters $t(i)$ and $p$.  

Using deductive reasoning based on the shaded cells of the matrix in Figure~\ref{fig:matrix}, one term of $N(i,1)$+$N(i,2)$, the number of participants of type $i$ who are assigned to the intervention group,  is equal to zero for each type $i$. Therefore, $N(1,2)$ is a binomial random variable with parameters $t(1)$ and $p$, $N(2,2)$ is a binomial random variable with parameters $t(2)$ and $p$, $N(3,1)$ is a binomial random variable with parameters $t(3)$ and $p$, and $N(4,1)$ is a binomial random variable with parameters $t(4)$ and $p$. Therefore, $N(1,2)$, $N(2,2)$, $N(3,1)$, and $N(4,1)$, the numbers of participants of each type assigned to the intervention group, are independent random variables with known binomial distributions.          

Furthermore, the numbers of participants of each type assigned to the control group, $N(1,4)$, $N(2,3)$, $N(3,4)$, and $N(4,3)$ are known functions of the numbers of each type assigned to the intervention group.  The number of participants of type $i$ in the control group, $N(i,3)+N(i,4)$ is equal to $t(i)$ minus the number of participants in the intervention group, $N(i,1)+N(i,2)$.  For example, the number of type $i=1$ participants in the control group, $N(1,4)$ is equal to the constant $t(1)$ minus $N(1,2)$, a binomial random variable with parameters $t(1)$ and $p$.  Therefore, it is possible to express all of the nonzero participant counts $N(i,j)$ in terms of four independent random variables with known binomial distributions, $N(1,2)$, $N(2,2)$, $N(3,1)$, and $N(4,1)$, and five constant terms: $t(1)$, $t(2)$, $t(3)$, and $t(4)$, and $p$.  For simplicity, I refer to the vector of the four constant terms $t(1)$, $t(2)$, $t(3)$, and $t(4)$ as the type vector $\mathbf{t}$.        

In each column of the matrix in Figure~\ref{fig:matrix}, let $G(j)$ denote the number of participants in group $j$, and denote a realization of $G(j)$ with $g(j)$.  The data consist of $G(1)$, $G(2)$, $G(3)$, and $G(4)$.  For simplicity, I refer to the vector of $G(1)$, $G(2)$, $G(3)$, and $G(4)$ with $\mathbf{G}$, and I denote a realization with $\mathbf{g}$.  With the notation, I emphasize that the data are grouped data.  Per the matrix, each group $j$ includes participants of two different types, and each $G(j)$ is the sum of two nonzero participant counts $N(i,j)$.  Because I can express all of the nonzero participant counts $N(i,j)$ in terms of the type vector $\mathbf{t}$ and the intended fraction of participants in the intervention group $p$, I can also express the probability of the data vector $\mathbf{G}$ as a function of the product of four independent binomial random variables and the constants $\mathbf{t}$ and $p$:     
\begin{align}
P(\mathbf{G}=\mathbf{g}\mid \mathbf{t},p) &= \sum_{\ell=0}^{t(1)} \mathrm{binom}\Big(\ell,t(1),p\Big) \nonumber \\
&\quad\;\ \times \mathrm{binom}\Big(g(2)-\ell,t(2),p\Big) \nonumber \\
&\quad\;\ \times \mathrm{binom}\Big(t(1)+t(3)-g(4)-\ell,t(3),p\Big) \nonumber \\
&\quad\;\ \times \mathrm{binom}\Big(g(1)+g(4)+\ell-t(1)-t(3),t(4),p\Big) \nonumber \\
&\quad\;\ \times \mathbf{1}\left\{\sum_{j=1}^{4}g(j)=\sum_{i=1}^{4}t(i)\right\}  \label{eq:dataprob}
\end{align}
where $\mathbf{1}\{\cdot\}$ is the indicator function and $\mathrm{binom}(\cdot)$ is the binomial probability mass function,
\begin{align*}
\mathrm{binom}(k,r,p) = P(K=k\mid r,p) = \binom{r}{k} p^k (1-p)^{r-k}.
\end{align*}
For completeness, I include the steps of the derivation in Appendix~\ref{appendix:proof}.

\section{Estimation}\label{sec:estimation}
\subsection{Maximum Likelihood Estimator} \label{sec:mle}

Equation \ref{eq:dataprob} also yields the likelihood of the type vector $\mathbf{t}$, given the realized data vector $\mathbf{g}$ and the intended fraction of participants in the intervention group $p$. It also yields the the likelihood of the type vector $\mathbf{t}$ and the intended fraction of participants in the intervention group $p$, given the realized data vector $\mathbf{g}$:        
\begin{equation*}
P(\mathbf{G}=\mathbf{g} \mid \mathbf{t},p)= \mathcal{L}(\mathbf{t} \mid \mathbf{g},p) = \mathcal{L}(\mathbf{t},p \mid \mathbf{g}). 
\end{equation*}

\noindent Therefore, the following is a maximum likelihood estimator for the type vector $\mathbf{t}$: \begin{equation}
\max_{\mathbf{t} \in \mathbb{N}_{0}^4} \; \mathcal{L}(\mathbf{t} \mid \mathbf{g},p) \nonumber 
\end{equation} 
where $\mathbf{t} \in \mathbb{N}_{0}^{4}$ indicates that the four elements of the type vector must belong to the set that includes the natural numbers and zero, which follows because they represent the counts of never takers, defiers, compliers, and always takers.

I propose the following alternative maximum likelihood estimator that does not require the intended fraction of participants in the intervention group $p$ to be specified: 
\begin{equation}
\max_{\mathbf{t} \in \mathbb{N}_{0}^4 , p \in (0,1)} \; \mathcal{L}(\mathbf{t}, p \mid \mathbf{g}) \nonumber, 
\end{equation} 
where $p \in (0,1)$ indicates that the intended fraction of participants in the intervention group $p$ must be a fraction between 0 and 1. Both maximum likelihood estimators are nonlinear programming problems that are difficult to solve, but a least squares estimator that also follows from the model is more computationally tractable.      
  
\subsection{Least Squares Estimator} \label{sec:ls}

Let $I\subseteq\{1,2,3,4\}$ represent any subset of the elements of the type vector $\mathbf{t}$. Using this notation, the randomization error in any subset of the type vector $\mathbf{t}$ is as follows:
\begin{equation*}
\varepsilon(I) = \sum_{i\in I}\left[N(i,1)+N(i,2)\right] - p\sum_{i\in I}t(i). \label{eq:err}
\end{equation*}

\noindent The randomization error represents the difference between the actual and intended number of participants in the intervention group.  The sum of the squared randomization error in all possible subsets of the type vector $\mathbf{t}$, weighted by the variance of the randomization error in each subset, yields the following objective function $\mathcal{S}$:
\begin{equation*}
\mathcal{S}(\mathbf{N}\mid p) = \sum_{I} \frac{1}{p(1-p)\sum_{i\in I} t(i)}\varepsilon(I)^2, \label{eq:leastSquares}
\end{equation*}
where $\mathbf{N}$ denotes the matrix of all $N(i,j)$.  Denote each element of the solution to the following nonlinear programming problem with $\hat{N}(i,j)$:
\begin{align*}
\min_{\mathbf{N} \in \mathbb{N}_{0}^{4\times 4}} &\; S(\mathbf{N}\mid p) \label{pr:ss}\\
\textrm{s.t.}
&\; \sum_{i=1}^{4} N(i,j) = g(j) \; \mathrm{for}\  j=1,2,3,4, \nonumber \\
&\; N(1,1)=N(1,3)=N(2,1)=N(2,4)=N(3,2)=N(3,3)=N(4,2)=N(4,4)=0 \nonumber 
\end{align*}
where $\mathbf{N} \in \mathbb{N}_{0}^{4\times 4}$ indicates that each element $N(i,j)$ of the 4 by 4 matrix $\mathbf{N}$ must belong to the set that includes the natural numbers and zero. The least squares estimate of each element of the type vector $\mathbf{t}$, which I denote with $\hat{t}(i)$, is as follows:
\begin{equation*}
\hat{t}(i) = \sum_{j=1}^{4}\hat{N}(i,j)\qquad \forall i = 1,2,3,4.
\end{equation*}

I also propose a separate version of the least squares estimator that does not require the intended fraction of participants in the intervention group $p$ to be specified.  That version of the estimator includes $p$ as a parameter to be estimated in the nonlinear programming problem, instead of taking $p$ as given.

\section{Empirical Application} \label{sec:empirical}

\subsection{Data}

I do not need access to microdata to apply the model to \citet{angrist1998} if I can obtain the grouped data that I need from a table that reports the cross-tabulation of the instrument and the treatment.  In the \citet{angrist1998} context of interest, the instrument is an indicator for whether a woman's first two children are of the same sex, and the treatment is an indicator for whether the woman has a third child.  Table 3 of \citet{angrist1998} reports a related cross-tabulation, but the results are in terms of fractions instead of counts.  To avoid rounding error in the implied counts, I derive the grouped data from publicly available microdata.\footnote{I obtain the data used by \citet{angrist2013} in their replication of \citet{angrist1998} from \url{http://sites.bu.edu/ivanf/files/2014/03/m_d_806.dta_.zip}. The note in Table 1 of \citet{angrist2013} states that the sample includes women aged 21--35 with at least two children.  The Stata code obtained from the same website as the data, which replicates the reported results, indicates additional restrictions to mothers who were at least 15 years old at the birth of their first child, whose oldest child is less than 18, and whose second child is older than one year.  The sample includes 394,840 observations. \citet{angrist1998} report that their comparable sample includes 394,835 observations.} The microdata include women in the 1980 US Census 5 percent Public Use Microdata Sample aged 21--35, who were at least 15 years old at the birth of their first child, who have two or more children, of which the oldest child is less than 18 and the second child is older than one year.  

\begin{figure}[!htb]
	\caption{Estimates from Replication of \citet{angrist1998}}
	\centering
	\includegraphics[width=\textwidth]{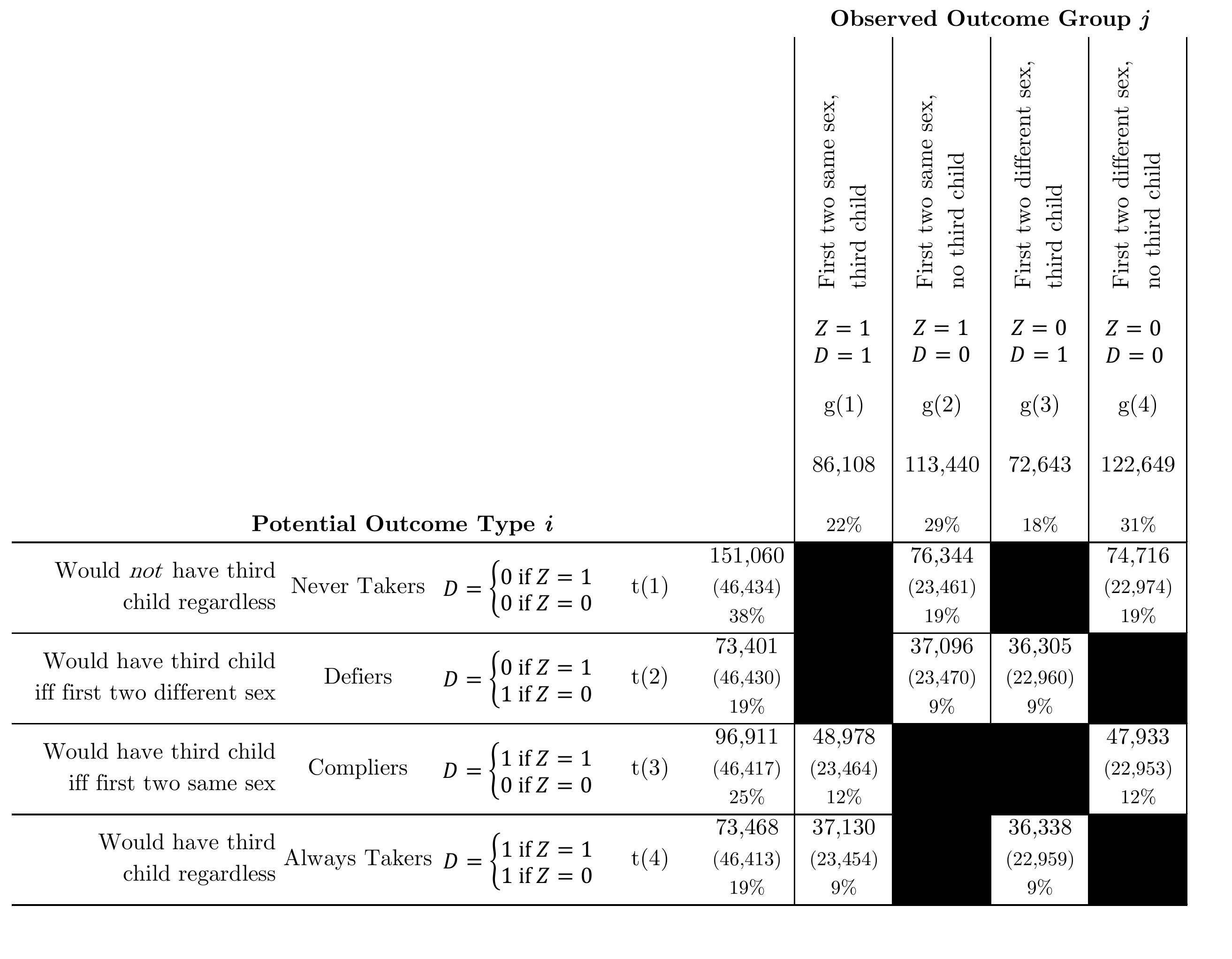}
	\label{fig:ptEstFstStg}
	\vspace{-20pt}
	\begin{minipage}{1\linewidth}
	\scriptsize
	\emph{Note}. $D$ is an indicator for whether the women has a third child, and $Z$ is an indicator for whether the woman's first two children are of the same sex. $N(i,j)$, the number of women of potential outcome type $i$ in observed outcome group $j$, must be equal to zero in all shaded cells. The observed outcome group vector $\mathbf{g}$ is calculated from a sample that includes women in the 1980 US Census 5 percent Public Use Microdata Sample aged 21--35, who were at least 15 years old at the birth of their first child, who have two or more children, of which the oldest child is less than 18 and the second child is older than one year. I estimate the intended fraction of participants in the intervention group $p$ as the empirical fraction of participants in the intervention group:  $p\approx 0.51$. I obtain estimates of the values of the matrix $\mathbf{N}$ and the potential outcome type vector $\mathbf{t}$ via the least squares estimator using the BARON \citep{baron} optimization package within MATLAB \citep{matlab2016}. Parentheses contain standard errors obtained as the standard deviation of estimates from 1000 bootstrap samples.  Percentages represent percentages of the full sample.  They do not necessarily sum to 100\% due to rounding.     
	\end{minipage}
\end{figure}

I report the required grouped data in the columns of Figure~\ref{fig:ptEstFstStg}.  There are $g(1)=86,108$ women who whose first two children are of the same sex who have a third child, and there are $g(2)=113,440$ women whose first two children are of the same sex who do not have a third child.  There are $g(3)=72,643$ women whose first two children are of different sex who have a third child, and there are $g(4)=112,649$ women whose first two children are of different sex who do not have a third child.   

\subsection{Replication}

Using the grouped data, I replicate the first stage estimate of \citet{angrist1998}.  The first stage estimate is equal to the fraction of the intervention group that is treated minus the fraction of the control group that is treated.  The first stage estimate indicates that mothers whose first two children are of the same sex are approximately 6 percentage points more likely to have a third child ($(86,108/(86,108+113,440) - 72,643/(72,643+122,649) \approx 0.06 $).  

Under LATE monotonicity, the first stage estimate of the impact of the instrument on the treatment gives the share of compliers \citep{angrist1996}. Therefore, compliers make up approximately 6 percent of the sample under LATE monotonicity.  In the \citet{angrist1998} context, compliers are mothers who would have a third child if and only if their first two children were of the same sex.  In addition to compliers, LATE monotonicity also allows for never takers and always takers but not defiers.  I include descriptions of all four types of women in the \citet{angrist1998} context in the rows of the matrix in Figure~\ref{fig:ptEstFstStg}.

Never takers are women who would \emph{not} have a third child regardless of the sex mix of their first two children. Under LATE monotonicity, there are no defiers, so we can deduce that women with two children of the same sex and no third child must be never takers.  We cannot identify all of the never takers in the sample because women with two children of different sex and no third children could be never takers or compliers.  However, under LATE independence, the expected share of always takers with two children of the same sex should be the same as the expected share of always takers in the sample. Therefore, as shown by \citet{imbens1997}, we can use the grouped data to estimate that always takers make up approximately 37 percent of the sample under LATE monotonicity ($72,643/(72,643+122,649) \approx 0.37 $).   

Always takers are women who would have a third child regardless of the sex mix of their first two children.  Under LATE monotonicity, there are no defiers, so we can deduce that women with two children of different sex and a third child must be always takers.  We cannot identify all of the always takers in the sample because women with two children of the same sex and a third child could be always takers or compliers.  However, under LATE independence, as shown by \citet{imbens1997}, we can estimate that the share of always takers in the full sample is equal to the share of always takers with two children of different sex.  Therefore, we can use the grouped data to estimate that never takers make up approximately percent of the sample under LATE monotonicity ($113,440/(113,440+86,108) \approx 0.57$).     

\subsection{Results}
The maximum likelihood estimator is not empirically tractable in this application via a variety of approaches, but I report the estimates that I obtain via the least squares estimator in Figure~\ref{fig:ptEstFstStg}.  Because the application is a natural experiment, I use the version of the least squares estimator that estimates the intended fraction of participants in the intervention group $p$.  In practice, when I estimate $p$ along with the other parameters, I obtain an estimate of $p$ that is equal to the empirical fraction of participants in the intervention group within reported precision, which is approximately 51\%.  Therefore, in the reported results, to increase computational efficiency, I  first estimate $p$ as the empirical fraction of participants in the intervention group, and then I estimate the other parameters.

I find that the sample includes 73,401 defiers, which represent approximately 19\% of the sample.  I also find that the sample includes 96,911 compliers, which represent approximately 25\% of the sample. In general, the first stage estimate is equal to the share of compliers minus the share of defiers \citep{angrist1996}.  Indeed, the first stage estimate of 6\% is equal to 25\% minus 19\%.  Under LATE monotonicity, the first stage estimate is an estimate of the share of compliers because the share of defiers is zero.  In this context, because the share of defiers is so large, the first stage estimate of 6\%  is a dramatic underestimate of the 25\% share of compliers.   

Defiers and compliers both respond to the sex mix of the first two children in their subsequent fertility. Together, defiers and compliers represent approximately 44\% of the sample, but they only make up 6\% of the sample under LATE monotonicity.  Therefore, subsequent fertility is much more responsive to the sex mix of the first two children when defiers are allowed.

Among the types that do not respond to the sex mix of the first two children, the never takers and the always takers, I find that the estimated shares under LATE monotonicity overestimate the size of both groups by approximately the same amount. I find that the sample includes 151,060 never takers, which represent approximately 38\% of the sample, as opposed to 57\% under LATE monotonicity.  I also find that the sample includes 73,468 always, which represent approximately 19\% of the sample, as compared to 37\% under LATE monotonicity.   

\section{Discussion}\label{sec:discussion}

The presence of defiers complicates the interpretation of the instrumental variable estimate.  Under LATE monotonicity, the instrumental variable estimate is the average treatment effect on compliers.  In the absence of LATE monotonicity, the instrumental variable estimate is equal to a weighted average of the treatment effect on compliers and the treatment effect on defiers \citep{angrist1996}. Even the reduced form estimate is difficult to interpret in the presence of defiers.  The textbook \emph{Mostly Harmless Econometrics} notes that ``We might therefore have a scenario where treatment effects are positive for everyone yet the reduced form is zero because effects on compliers are canceled out by effects on defiers" (\citet{angrist2009} page 156). 
 
If the treatment effects on compliers and defiers are the same, then the instrumental variable estimate still gives the treatment effect on compliers \citep{angrist1996}.  In the model that eliminates LATE monotonicity, I could therefore impose an ancillary assumption that treatment effects on compliers and defiers are the same to restore the interpretability of the LATE.  However, one important advantage of the LATE framework is that it allows for treatment effect heterogeneity, so the assumption that treatment effects are the same for compliers and defiers is not necessarily desirable in general.  For now, estimates of counts of defiers can inform the plausibility of various ancillary assumptions on a case-by-case basis.  

Estimates of the counts of defiers can also inform the plausibility of the LATE independence assumption.  In the context of \citet{angrist1998}, the presence of a large share of defiers is plausible.  In other contexts, the presence of a large share of defiers might not be plausible.  In those contexts, an estimate of a large share of defiers could indicate a violation of the LATE independence assumption.  In the context of a randomized experiment that is implemented carefully, the LATE independence assumption should be plausible.

\section{Conclusion}\label{sec:conclusion}

The LATE framework of \citet{imbens1994} imposes an asymmetry between the first stage and the reduced form via an assumption of monotonicity in the first stage.  In this paper, I eliminate the monotonicity assumption in the first stage, which allows for defiers.  I estimate the count of defiers in an empirical application to \citet{angrist1998}, and I show that defiers represent a meaningful share of the sample.  My estimates call into question the validity of the LATE monotonicity assumption in that context.

\appendix

\setcounter{equation}{0}\renewcommand\theequation{A\arabic{equation}} 

\section{Appendix: Derivation of Equation \ref{eq:dataprob}}\label{appendix:proof}

Below, I transition from (\ref{eq:start}) to (\ref{eq:eight}) by expressing each element of the data vector as the sum of two numbers of participants using deductive reasoning per the matrix in Figure~\ref{fig:matrix}.  Next, I transition to (\ref{eq:four}) by expressing all terms as functions of the numbers of participants of each type assigned to the intervention group, $N(1,2)$, $N(2,2)$, $N(3,1)$, and $N(4,1)$, which are independently distributed binomial random variables.  Because each term of the joint probability in (\ref{eq:four}) represents the sum of two independently distributed binomial random variables, I apply the approach to deconvolution for discrete random variables to transition to (\ref{eq:decon}).  I rearrange terms in (\ref{eq:deconsingle}) so that each term of the joint probability represents a single binomial random variable.  Finally, because all terms in the resulting joint probability are independent, I express the joint probability as the product of four independent probabilities. 
\begin{align}
P(\mathbf{G}=\mathbf{g} \mid \mathbf{t},p) &= P\bigg(G(1)=g(1), G(2)=g(2),G(3)=g(3),G(4)=g(4) \mid \mathbf{t},p\bigg) \label{eq:start} \\
&= P\bigg(N(3,1)+N(4,1)=g(1), \nonumber \\ 
&\qquad\quad N(1,2)+N(2,2)=g(2), \nonumber \\
&\qquad\quad N(2,3)+N(4,3)=g(3),  \nonumber \\
&\qquad\quad N(1,4)+N(3,4)=g(4) \mid \mathbf{t},p\bigg) \label{eq:eight} \\
&= P\bigg(N(3,1)+N(4,1)=g(1), \nonumber \\
&\qquad\quad N(1,2)+N(2,2)=g(2), \nonumber \\
&\qquad\quad N(2,2)+N(4,1)=t(2)+t(4)-g(3), \nonumber \\
&\qquad\quad N(1,2)+N(3,1)=t(1)+t(3)-g(4) \mid \mathbf{t},p\bigg) \label{eq:four} \\
&= \sum_{\ell=0}^{t(1)} P\bigg(N(1,2)=\ell, \nonumber \\
&\qquad\qquad\;\ N(3,1) + N(4,1)=g(1), \nonumber \\
&\qquad\qquad\;\ \ell+N(2,2)=g(2), \nonumber \\
&\qquad\qquad\;\ N(2,2)+N(4,1)=t(2)+t(4)-g(3), \nonumber \\
&\qquad\qquad\;\ \ell+N(3,4)=t(1)+t(3)-g(4) \mid \mathbf{t},p\bigg) \label{eq:decon} \\
&= \sum_{\ell=0}^{t(1)} P\bigg(N(1,2)=\ell, \nonumber \\
&\qquad\qquad\;\ N(2,2)=g(2)-\ell,\nonumber  \\
&\qquad\qquad\;\ N(3,1)=t(1)+t(3)-g(4)-\ell, \nonumber \\
&\qquad\qquad\;\ N(4,1)=g(1)+g(4)+\ell-t(1)-t(3), \nonumber \\
&\qquad\qquad\;\ \sum_{j=1}^4 g(j)=\sum_{i=1}^4 t(i) \mid \mathbf{t},p\bigg) \label{eq:deconsingle} \\
&= \sum_{\ell=0}^{t(1)} P\Big(N(1,2)=\ell \mid \mathbf{t},p\Big) \nonumber \\
&\quad\;\ \times P\Big(N(2,2)=g(2)-\ell \mid \mathbf{t},p\Big) \nonumber \\
&\quad\;\ \times P\Big(N(3,1)=t(1)+t(3)-g(4)-\ell \mid \mathbf{t},p\Big) \nonumber \\
&\quad\;\ \times P\Big(N(4,1)=g(1)+g(4)+\ell-t(1)-t(3) \mid \mathbf{t},p\Big) \nonumber \\
&\quad\;\ \times \mathbf{1}\left\{\sum_{j=1}^4 g(j) = \sum_{i=1}^{4}t(i)\right\} \label{eq:penultimate} \\
&= \sum_{\ell=0}^{t(1)} \mathrm{binom}\Big(\ell,t(1),p\Big) \nonumber \\
&\quad\;\ \times \mathrm{binom}\Big(g(2)-\ell,t(2),p\Big) \nonumber \\
&\quad\;\ \times \mathrm{binom}\Big(t(1)+t(3)-g(4)-\ell,t(3),p\Big) \nonumber \\
&\quad\;\ \times \mathrm{binom}\Big(g(1)+g(4)+\ell-t(1)-t(3),t(4),p\Big) \nonumber \\
\setcounter{equation}{0}\renewcommand\theequation{\arabic{equation}} 
&\quad\;\ \times \mathbf{1}\left\{\sum_{j=1}^{4}g(j)=\sum_{i=1}^{4}t(i)\right\} \nonumber
\end{align}
where $\mathbf{1}\{\cdot\}$ is the indicator function and $\mathrm{binom}(\cdot)$ is the binomial probability mass function,
\begin{align*}
\mathrm{binom}(k,r,p) = P(K=k\mid r,p) = \binom{r}{k} p^k (1-p)^{r-k}.
\end{align*}

\bibliographystyle{chicago}
\bibliography{fertility}

\end{document}